\documentclass[11pt,twoside]{article}
\usepackage{asp2010}
\usepackage{graphicx}

\resetcounters

\begin{document}

\title{Kiloparsec-Scale Simulations of Magnetised Molecular Clouds in Disk Galaxies.}
\author{Sven Van Loo,$^1$ Michael J. Butler,$^2$ Jonathan C. Tan,$^2$ and Sam A. E. G. Falle$^3$
\affil{$^1$Harvard-Smithsonian Center for Astrophysics, 60 Garden Street, Cambridge, MA 02138, USA}
\affil{$^2$Department of Astronomy, University of Florida, Gainesville, FL 32611, USA}
\affil{$^3$Department of Applied Mathematics, University of Leeds, Woodhouse Lane, Leeds, LS2 9JT, UK}}

\begin{abstract}
We present simulations of the evolution of self-gravitating dense gas 
on kiloparsec-size scales in a galactic disk, designed 
to study dense clump formation from giant molecular clouds (GMCs). These dense clumps 
are expected to be the precursors to star clusters and this process may be the rate
limiting step controling star formation rates in galactic systems as described by 
the Kennicutt-Schmidt relation. 
The evolution of these simulated GMCs and clumps is determined by self-gravity balanced 
by turbulent pressure support and the large scale galactic shear. 
While the cloud structures and densities significantly change during their evolution, they
remain roughly in virial equilibrium for time scales exceeding
the free-fall time of GMCs, indicating that energy from the galactic shear
continuously cascades down.  We implement star formation at a slow, 
inefficient rate of 2\% per local free-fall time, but this still yields global star 
formation rates that are $\sim$ two orders of magnitude larger than the observed
Kennicutt-Schmidt relation due to the over-production of dense clump gas. 
To explain this discrepancy, we anticipate magnetic fields to provide additional
support. Low-resolution simulations indeed show that the magnetic field 
reduces the star formation rate.
\end{abstract}

\section{Introduction}
Star formation in galaxies involves a vast range of length and time
scales, from the tens of kiloparsec diameters and $\sim 10^8$~yr
orbits of galactic disks to the $\sim 0.1$~pc sizes and $\sim 10^5$~yr
dynamical times of individual prestellar cores (PSCs)
\citep[see][for a review]{McKeeOstriker2007}. The interstellar medium 
(ISM) contains various forms of pressure support (including
thermal, magnetic and turbulent), large-scale coherent motions (including
Galactic shear and large-scale turbulent flows) that drive turbulent motions,
and localized feedback from newborn stars that effectively counter self-gravity.
The overall star formation rate is then relatively slow and inefficient at just a few percent
conversion of gas to stars per local dynamical timescale across a wide
range of densities \citep[][]{KrumholzTan2007}. However, the relative
importance of the above processes for suppressing star formation is
unknown, even for the case of our own Galaxy.

To investigate the star formation process within molecular clouds, a
significant range of the internal structure of GMCs needs to be resolved
including dense gas clumps expected to be the birth locations of star
clusters. 
The global galaxy simulations of \citet[][hereafter TT09]{Tasker&Tan2009} 
followed the formation and evolution of
thousands of GMCs in a Milky-Way-like disk with a flat rotation
curve. However, with a spatial resolution of $\sim$8~pc, only the
general, global properties of the GMCs could be studied; not their
internal structure. Following all of these GMCs to higher spatial
resolution is very expensive in terms of computational resources.
Therefore we need to use an alternative method.

Our approach is to extract a 1 kpc by 1 kpc patch of the disk,
extending 1~kpc both above and below the midplane, centered at a
radial distance of 4.25~kpc from the Galactic center.
This is done at a time 250~Myr after
the beginning of the TT09 simulation, when the disk has fragmented
into a relatively stable population of GMCs. We then follow the
evolution of the ISM, especially the GMCs, including their
interactions, internal dynamics and star formation activity, down to
clump-size, or parsec-size, scales. These local
simulations are able to reach higher densities, resolve smaller mass
scales and include extra physics compared to the global simulations.

The grid resolution of the initial conditions is 128$^2 \times$ 256
which corresponds to a cell size of 7.8\ pc and serves as the root
grid for the high resolution simulations. Most of the simulations we
present here involve 4 levels of adaptive mesh refinement of the root
grid, thus increasing the effective resolution to 2048$^2 \times 4096$
or about 0.5~pc. Refinement of a cell occurs when the Jeans length
drops below four cell widths, in accordance with the criteria
suggested by \citet[][]{Trueloveetal1997} for resolving gravitational
instabilities.
The simulations performed in this paper were run using {\it Enzo}
\citep[e.g.][]{OSheaetal2004}. 

\section{Resolving the GMCs}
We first carryout the simulation with the physics
and resolution identical to the global galaxy simulation of TT09,
i.e. a grid resolution of $\sim$ 8pc and only the \citet[][]{RosenBregman} cooling
function down to 300K is included.

\begin{figure}\label{fig:uni_amr}
\plottwo{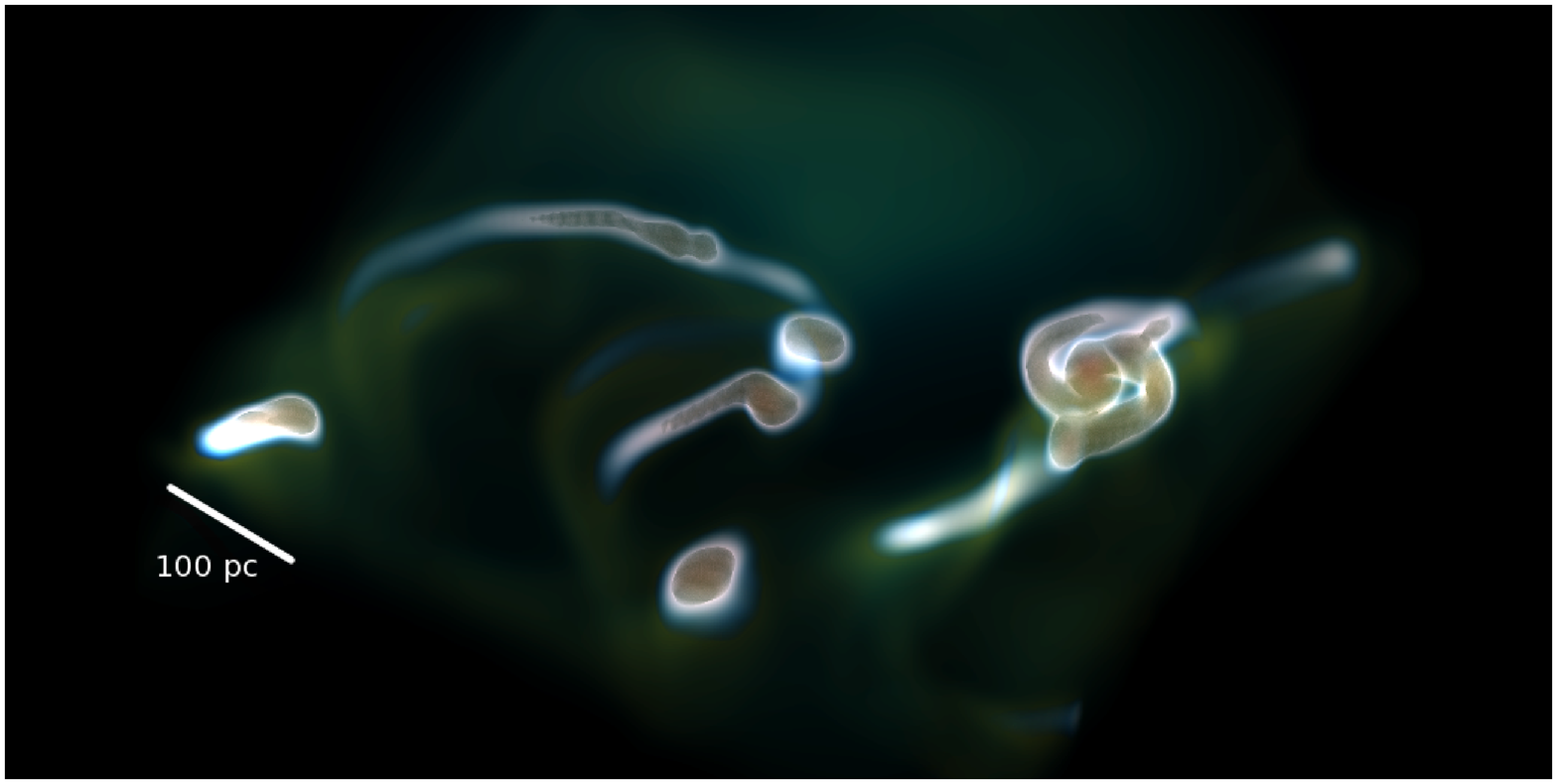}{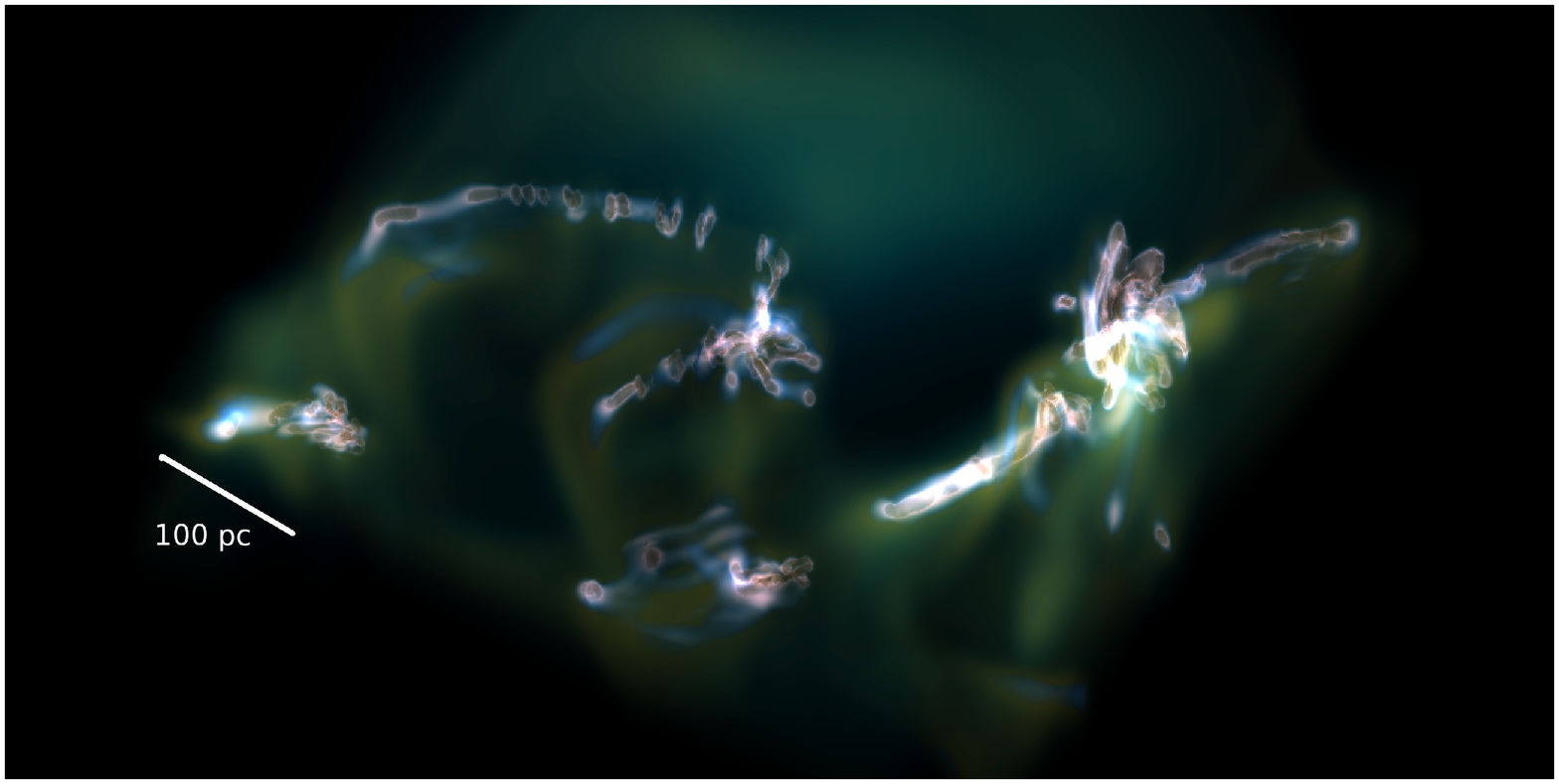}
\caption{Volume rendering of 
the number density after 10 Myr for a uniform (left) and AMR (right) 
simulation with the same physics as TT09. ``GMC'' gas is coloured blue, while 
``Clump'' gas is red. The white line shows a distance of 100 pc.}
\end{figure}

\begin{figure}\label{fig:molecular_mfrac}
\begin{center}
\includegraphics[width=0.52\textwidth]{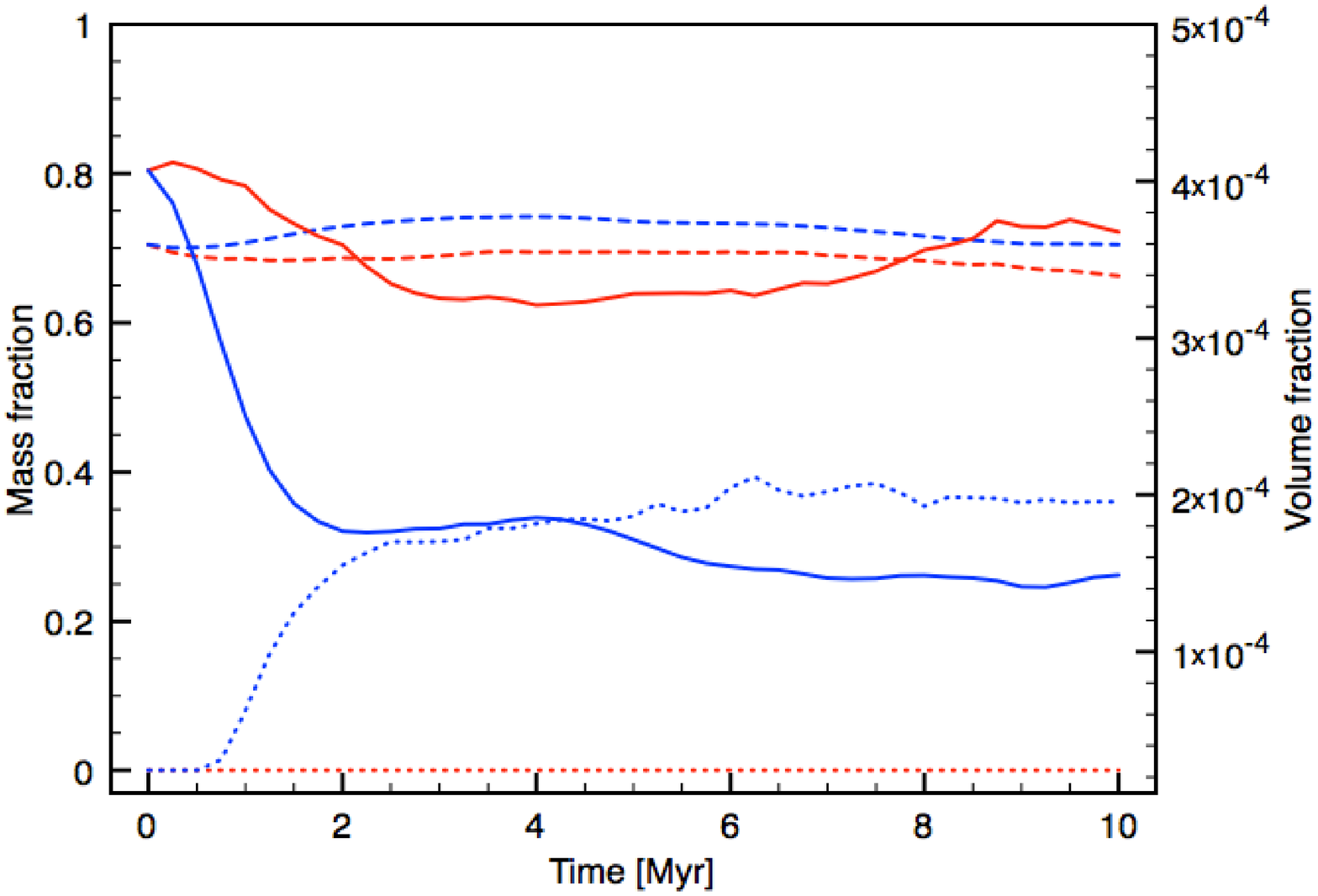}
\caption{The mass fraction of gas in ``GMCs'' 
(dashed) and in ``Clumps'' (dotted) and the volume fraction of
gas in ``GMCs'' (solid) for the uniform (red) and AMR model (blue).}
\end{center}
\end{figure}

Figure~1 shows the gas density for this simulation
(left panel) after 10\ Myr. While the clouds show some signs 
of gravitational contraction, interaction and fragmentation over this
time scale, they  have not changed dramatically. In fact, the mass 
fraction of ``GMC'' gas (i.e. $n_{\rm H} > 10^2~{\rm cm^{-3}}$) remains 
quite constant throughout the simulation, which spans a few free-fall times
(Fig.~\ref{fig:molecular_mfrac}). The volume fraction of ``GMC" gas 
shows an initial decrease, but also remains relatively constant. 
Thus the clouds are roughly in virial equilibrium and this is reflected by 
the virial parameters of the clouds which are between 0.73 and 1.1 
\citep[][]{VanLooetal2012}. The thermal pressure and 
non-thermal motions within the clouds provide enough support against 
self-gravity to prevent runaway gravitational collapse (even though 
the non-thermal motions are not well resolved in this uniform grid 
simulation).  The nonthermal motions are 
continuously driven by bulk cloud motions in a shearing galactic disk
as kinetic energy cascade down.

By increasing the resolution down to $\sim 0.5$\ pc the GMCs
can now be better resolved and the evolution of dense clumps
within the clouds begins to be captured. While the mass fraction of ``GMC'' gas increases 
slightly, the volume fraction decreases by a factor of 2 within 2\ Myr 
after which it remains roughly constant (see Fig.~\ref{fig:molecular_mfrac}). 
In roughly the same time span, about half of the ``GMC'' gas accumulates in ``Clumps''
(i.e. $n_{\rm H} > 10^5~{\rm cm^{-3}}$). 
Then the mass fraction in clumps remains nearly constant and reaches
a new quasi-steady state. So, initially, the gas within the clouds collapses
to form filaments with dense clumps due to the increased resolution (see right panel
of Fig.~1). Concurrent with the contraction, the velocity dispersion in the clouds 
increases. The virial parameters of the clouds now lie between 1.6 and 4.5
\citep[][]{VanLooetal2012}. Non-thermal 
motions again counter the effects of self-gravity to virialise the cloud. 
The initial evolution of the clouds is thus a direct consequence of the increase 
in resolution.

\section{Molecular cooling}
While the increased resolution helps to describe the substructures of
GMCs in greater detail, the thermal properties of the ISM are poorly
reproduced. As only cooling is included in the simulations of the previous section, 
most of the gas within the disk is at the floor temperature of 300\ K. To
reproduce the multiphase character of the ISM, i.e. a cold, dense
and a warm, diffuse phase, we include diffuse heating. Additionally,
we study the influence of different cooling functions.  While atomic
cooling can be adequately described by the Rosen \& Bregman cooling
function, the \citet[][]{SanchezSalcedoetal2002} function extends down 
to a temperature of 5\ K, i.e.  nearly two orders of magnitude lower, 
and includes a thermally unstable temperature range. 
To include molecular cooling, especially from H$_2$ and CO molecules, we adopt 
a novel approach. Using the simulation with the atomic cooling, the 
column extinction can be expressed as function of gas density. Such a 
one-to-one relation eliminates the need for time-consuming column-extinction 
calculations to assess the attenuation of the radiation field in the 
simulations. We also use this extinction-density relation to generate a table 
of cooling and heating rates as function of density and temperature with the 
code Cloudy \citep[][]{Ferlandetal1998}. More details on this approach are
given in \citet[][]{VanLooetal2012}.

Including diffuse heating mostly affects the gas outside the GMCs as 
this gas has a higher equilibrium temperature than 300\ K (i.e. the 
temperature of the gas in the disk in the initial simulations). Then 
a higher external pressure acts as an additional force confining 
the GMCs. The higher external pressure resulting from this heating 
of the disk is, however, much smaller than the internal pressure of the 
GMCs, which is set by their self-gravitating weight. 

\begin{figure}[!ht]\label{fig:PDF_multiphase}
\begin{center}
\includegraphics[width=0.52\textwidth]{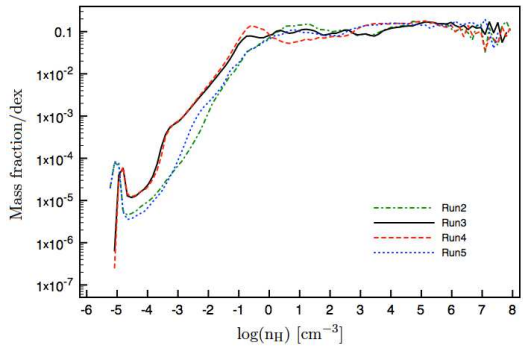}
\caption{Mass-weighted Probability Density Function
(PDF) for the Rosen \& Bregman cooling with (Run 3) and without heating (Run 2), 
Sanchez-Salcedo et al. (Run 4) and Cloudy cooling function (Run 5) simulations 
after 10\ Myr.}
\end{center}
\end{figure}

For the Sanchez-Salcedo et al. and Cloudy cooling function, the internal thermal
pressure decreases by $\sim$ 2 orders of magnitude (the floor temperature goes from 
300K to 5K). However, the global properties of the clouds remain similar, e.g.
the distribution of gas as function of density in these models is nearly identical
(see Fig.~\ref{fig:PDF_multiphase}). Thus, thermal pressure is negligible compared 
to self-gravity and turbulent pressure in setting the cloud properties.
The PDFs also reveal that the clouds evolve to a state where a large fraction,
30-40\% of the gas is in clumps. As in nearby GMCs this value is only between
0 and 15\% \citep[][]{Edenetal2012}, so there is a clear overproduction of dense 
clump gas in our simulations. The high clump formation efficiency is partly due 
to our resolution limit.  We do not properly capture the formation of individual 
clumps so that the turbulent dissipation range is not fully resolved. 
The clumps then lack turbulent support against self-gravity hereby attaining 
higher densities and accumulating more mass.

\begin{figure}[!ht]\label{fig:sfr}
\begin{center}
\includegraphics[width=0.52\textwidth]{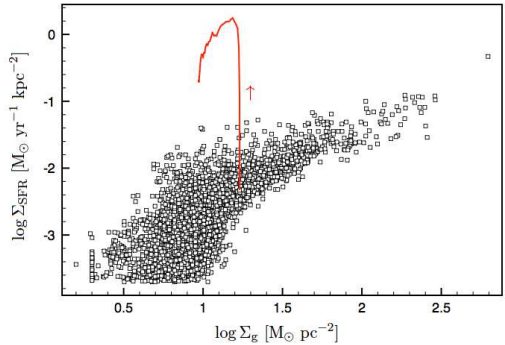}
\caption{The star formation rate ($\Sigma_{\rm SFR}$) as function
of the gas surface density. The symbols show the observations of \citet[][]{Bigieletal2008},
while the solid line shows the evolution of $\Sigma_{\rm SFR}$ in our simulation. The 
arrow shows the direction of the evolution.}
\end{center}
\end{figure}

\section{Star formation}
To model star formation, we allow collisionless star cluster
particles, representing a {\it star cluster} of mass $M_*$, to form in 
the dense clumps of our simulations. These star
cluster particles are created when the density within a cell exceeds a
fiducial star formation threshold value of $n_{\rm H,sf} =
10^5$\ cm$^{-3}$. We use a prescription of a fixed star formation efficiency per local
free-fall time, $\epsilon_{\rm ff}$, for those regions with $n_{\rm
  H}>n_{\rm H,sf}$.  Our fiducial choice of $\epsilon_{\rm ff}$ is 0.02 as  
implied by observational studies 
\citep[][]{ZuckermanEvans1974, KrumholzTan2007}.

When a cell reaches the threshold density, a star cluster particle is
created whose mass is calculated by
\begin{equation}\label{eq:mstar}
 M_* = \epsilon_{\rm ff} \frac{\rho \Delta x^3}{t_{\rm ff}} \Delta t,
\end{equation}
where $\rho$ is the gas density, $\Delta x^3$ the cell volume, $\Delta
t$ the numerical time step, and $t_{\rm ff}$ the free-fall time of gas in
the cell (evaluated as $t_{\rm ff} = (3\pi/32G\rho)^{1/2}$). To avoid too
many star particles of low mass, only star particles with a mass of 
$M_{\rm *, min}$=100$M_{\sun}$
are created with a probability of $M_*/M_{\rm *,min}$.
The motions of the star cluster particles are calculated as a
collisionless N-body system within the total potential due to galaxy, gas and stars. 

Star formation has not changed much of the global density structure or
dynamics, but has reduced the maximum gas surface densities by about
an order of magnitude compared to the model without star formation.
As the star formation rate depends on the gas mass in dense clumps 
(Eq.~\ref{eq:mstar}), the evolution of the star formation rate follows 
the clump mass fraction. It is then not surprising that 
we find mean star formation rate of 0.8\ M$_{\sun}$ yr$^{-1}$ kpc$^{-2}$
over the 10 Myr time scale. The star formation rate in our simulations 
exceeds that expected from the Kennicutt-Schmidt relation \citep[e.g.][]{Bigieletal2008}
by $\sim$ two orders of magnitude (see Fig.~4).
This over efficiency of star formation in our simulations
is a simple reflection of the high mass fraction in dense clumps.

\begin{figure}[!ht]\label{fig:MHD}
\begin{center}
\includegraphics[width=0.52\textwidth]{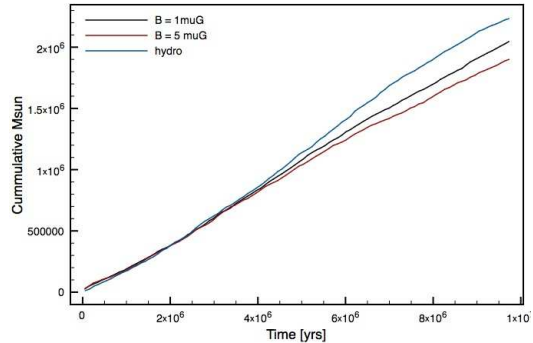}
\caption {Cumulative mass in star particles for 
uniform HD and MHD simulations.}
\end{center}
\end{figure}

\section{Magnetic fields}
Magnetic fields have been observed on large-scales in galaxies 
\citep[][]{Fletcher2010} down to small-scales in dense cores \citep[][]{Girart2009}.
They provide additional support to the GMCs and can play a role in reducing the 
high clump mass fraction found in the hydrodynamical simulations. 

To study the effect of magnetic fields, we ran MHD simulations
where the initial conditions are threaded with a uniform 
magnetic field along the $y$-axis (i.e. the direction of the shear). 
The magnitude of the field varies between 1 and 10 $\mu$G representing 
values around the mean Galactic value of $\sim 6 \mu$G. 
The uniform simulations show slightly expanding clouds due to the additional support
against self-gravity and less fragmentation with increasing magnetic field 
strength.  When we include the star formation routine, albeit with a lower 
$n_{\rm H,sf} = 10^2~{\rm cm^{-3}}$ and higher $M_* = 1000~{\rm M_{\sun}}$, fewer 
star particles form in the MHD simulations than in the hydro simulations 
(see Fig.~5). 
However, to understand the impact of the field on clump formation, higher 
resolution simulations are required (and are currently in progress).

\acknowledgements We thank Sam Skillman for 
the scripts to produce the rendered images. SvL acknowledges support 
from the SMA Postdoctoral Fellowship of
the Smithsonian Astrophysical Observatory and JCT 
from NSF CAREER grant AST-0645412; NASA Astrophysics Theory and
Fundamental Physics grant ATP09-0094; NASA Astrophysics Data Analysis
Program ADAP10-0110. Resources supporting this work were provided by
the NASA High-End Computing (HEC) Program through the NASA Advanced
Supercomputing (NAS) Division at Ames Research Center.

\end{document}